\documentclass[pdflatex,sn-mathphys-num]{sn-jnl}%
\usepackage{amsmath,amssymb,amsfonts}
\usepackage{graphicx}
\usepackage{booktabs}
\usepackage{array}
\usepackage{tabularx}
\usepackage{mathtools}
\usepackage{url}
\usepackage{tikz}
\usetikzlibrary{arrows.meta,positioning,fit}
\usepackage{listings}
\newcommand{\N}{\mathcal{N}}
\newcommand{\E}{\text{E}}
\newcommand{\Var}{\operatorname{Var}}

\newcommand{\GIG}{\operatorname{GIG}}
\newcommand{\Ga}{\operatorname{gamma}}

\newcommand{\trans}{^{\mathsf T}}
\begin{document}
\title[Deep Mixture Models]{Deep Mixture Models: A Unified Statistical Review with R}
\author[]{\fnm{Jinran} \sur{Wu}}\email{jinran.wu@uq.edu.au}
\author*[]{\fnm{Geoffrey J.} \sur{McLachlan}}\email{g.mclachlan@uq.edu.au}
\affil[]{ \orgname{The University of Queensland}, \orgaddress{\country{Australia}}}

\abstract{Deep mixture models extend finite mixture models through hierarchical latent factor structures with parameters shared across mixture pathways. This focused review examines deep Gaussian mixture models (DGMMs), robust deep mixture models (RDMMs), deep skew-$t$ mixture models (DStMMs), and mixed deep Gaussian mixture models (MDGMMs) within a common statistical framework. We distinguish pathway distributions, measurement models, and latent architectures, and relate these features to likelihood evaluation and parameter estimation. For continuous measurements, we connect Gaussian conditional calculations to the scale-mixture representations of Student-$t$ and skew-$t$ pathways. For mixed measurements, response-specific likelihoods introduce numerical integration over latent variables. The accompanying \texttt{MixedDeepMM} R package provides a unified interface for Gaussian, Student-$t$, and skew-$t$ kernels, continuous and mixed measurements, and single- and two-head architectures. Applications to the UCI Wine and Cryotherapy datasets illustrate the relationships among model fit, clustering agreement, and internal cluster separation through repeated fits and consensus partitions. The review concludes with computational and statistical issues arising from model complexity, architecture selection, and uncertainty assessment.}

\maketitle

\section{Introduction}\label{sec:intro}

Finite mixture models are widely used for density estimation, classification, and modelling population heterogeneity~\citep{TitteringtonSmithMakov1985, McLachlanPeel2000, FraleyRaftery2002, mclachlan2019finite}. In high dimensions, however, unrestricted component covariances can be costly: a Gaussian mixture with $k$ components and $p$ variables requires $kp(p+1)/2$ covariance parameters alone. Parsimonious covariance models and mixtures of factor analyzers (MFAs) address this problem by representing within-component dependence through lower-dimensional latent factors~\citep{CeleuxGovaert1995, GhahramaniHinton1997, McLachlanPeelBean2003, BaekMcLachlanFlack2010}.

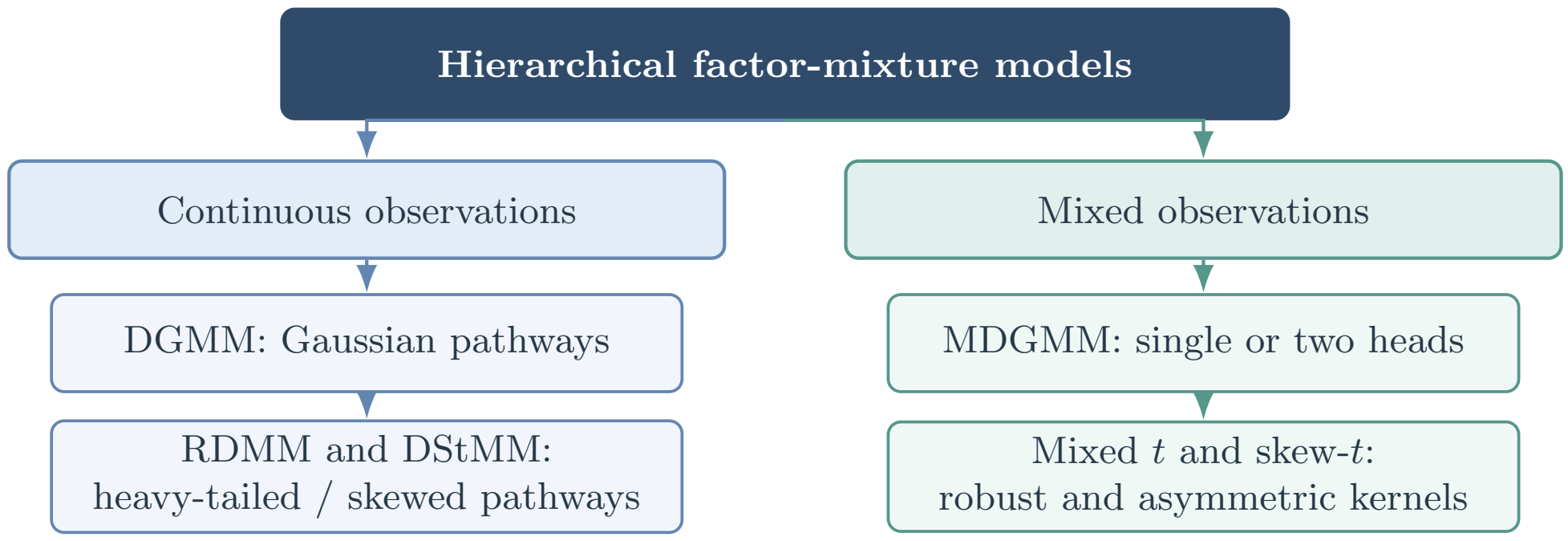
\begin{figure}[!ht]
\centering
\definecolor{RootBlue}{HTML}{304B6A}
\definecolor{ContBlue}{HTML}{E3EDF8}
\definecolor{ContLight}{HTML}{F2F6FC}
\definecolor{ContBorder}{HTML}{6286B1}
\definecolor{MixedTeal}{HTML}{E1F0ED}
\definecolor{MixedLight}{HTML}{F0F8F6}
\definecolor{MixedBorder}{HTML}{57968B}
\definecolor{TextDark}{HTML}{263747}
\begin{tikzpicture}[
    >=Latex,
    root/.style={
        draw=RootBlue,
        fill=RootBlue,
        text=white,
        rounded corners=3pt,
        align=center,
        minimum height=9mm,
        font=\small\bfseries,
        line width=0.8pt
    },
    cont/.style={
        draw=ContBorder,
        fill=ContBlue,
        text=TextDark,
        rounded corners=3pt,
        align=center,
        minimum height=8mm,
        font=\small,
        line width=0.8pt
    },
    mixed/.style={
        draw=MixedBorder,
        fill=MixedTeal,
        text=TextDark,
        rounded corners=3pt,
        align=center,
        minimum height=8mm,
        font=\small,
        line width=0.8pt
    },
    contleaf/.style={
        draw=ContBorder,
        fill=ContLight,
        text=TextDark,
        rounded corners=3pt,
        align=center,
        minimum height=8mm,
        font=\small,
        line width=0.7pt
    },
    mixedleaf/.style={
        draw=MixedBorder,
        fill=MixedLight,
        text=TextDark,
        rounded corners=3pt,
        align=center,
        minimum height=8mm,
        font=\small,
        line width=0.7pt
    },
    arrow/.style={-Latex,thick}
]

\node[root,minimum width=83mm] (root) at (0,2.9)
    {Hierarchical factor-mixture models};

\node[cont,minimum width=59mm] (cont) at (-3.45,1.7)
    {Continuous observations};

\node[mixed,minimum width=59mm] (mixed) at (3.45,1.7)
    {Mixed observations};

\node[contleaf,minimum width=52mm] (gau) at (-3.45,.6)
    {DGMM: Gaussian pathways};

\node[contleaf,minimum width=52mm] (heavy) at (-3.45,-.5)
    {RDMM and DStMM:\\heavy-tailed / skewed pathways};

\node[mixedleaf,minimum width=52mm] (heads) at (3.45,.6)
    {MDGMM: single or two heads};

\node[mixedleaf,minimum width=52mm] (extension) at (3.45,-.5)
    {Mixed $t$ and skew-$t$:\\robust and asymmetric kernels};

\draw[arrow,draw=ContBorder]
    (root.south) -| (cont.north);
\draw[arrow,draw=MixedBorder]
    (root.south) -| (mixed.north);

\draw[arrow,draw=ContBorder]
    (cont.south) -- (gau.north);
\draw[arrow,draw=ContBorder]
    (gau.south) -- (heavy.north);

\draw[arrow,draw=MixedBorder]
    (mixed.south) -- (heads.north);
\draw[arrow,draw=MixedBorder]
    (heads.south) -- (extension.north);

\end{tikzpicture}
\caption{Schematic relationships among the reviewed deep mixture specifications. Pathway kernels govern tail behaviour and asymmetry; mixed-response formulations also distinguish the common-head and two-head latent architectures.}
\label{fig:taxonomy}
\end{figure}

Hierarchical factor mixtures extend MFAs by introducing successive mixture layers between the observations and deeper latent factors. Deep mixtures of factor analyzers and deep Gaussian mixture models (DGMMs) yield flexible marginal densities while sharing layer-specific parameters across latent pathways~\citep{TangSalakhutdinovHinton2012, ViroliMcLachlan2019}. The hierarchy supports density estimation, clustering, and lower-dimensional representation; variational and sparse extensions address high-dimensional applications~\citep{KockKleinNott2022}. Two directions broaden this framework. For continuous data, robust deep mixture models (RDMMs) introduce a shared scale variable to produce Student-$t$ pathways, while deep skew-$t$ mixture models (DStMMs) also accommodate asymmetry~\citep{PeelMcLachlan2000, WuMcLachlan2026Robust, WuWangMcLachlan2026SkewT}. For heterogeneous data, mixed deep Gaussian mixture models (MDGMMs) combine latent Gaussian factors with measurement distributions for binary, ordinal, nominal, and count responses, using either a common latent head or separate continuous and discrete heads~\citep{MoustakiKnott2000, GolliniMurphy2014, McParlandGormley2016, FuchsPommeretViroli2022}. Although these models share a factor-mixture foundation, they differ in pathway distributions, observation models, and computational requirements. Figure~\ref{fig:taxonomy} provides a schematic overview of these model families, highlighting their distinctions in response types, pathway kernels, and latent architectures.

This focused review compares DGMMs, RDMMs, DStMMs, and MDGMMs in terms of latent pathways, observed likelihoods, and parameter estimation. We organise these models by response type, pathway kernel, and latent architecture within a single statistical formulation, linking their distributional assumptions to their inferential requirements. The accompanying \texttt{MixedDeepMM} R package~\citep{MixedDeepMM2026} provides a common interface for these specifications, illustrated using the UCI Wine and Cryotherapy datasets. Its main functions are summarised in Appendix~\ref{app:r}.

The remaining parts are organised as follows. Section~\ref{sec:models} introduces the model families, Section~\ref{sec:computing} develops the common likelihood and inferential framework, and Section~\ref{sec:applications} presents the data illustrations. Section~\ref{sec:discussion} discusses their connections and remaining challenges.

\section{Established deep mixture models}\label{sec:models}

\subsection{Gaussian factor-mixture hierarchies}\label{sec:gaussian}

Let $\boldsymbol y_i\in\mathbb R^p$, $i=1,\ldots,n$, denote independent observations. An MFA has density
\begin{equation}\label{eq:mfa}
 f(\boldsymbol y_i)=\sum_{a=1}^{k}\pi_a\,\phi_p\left(\boldsymbol y_i;\boldsymbol\eta_a,\boldsymbol\Lambda_a\boldsymbol\Lambda_a\trans+\boldsymbol\Psi_a\right),\quad
 \pi_a>0,\quad\sum_{a=1}^k\pi_a=1.
\end{equation}
Here $\boldsymbol\Lambda_a$ is a $p\times q$ loading matrix with $q<p$, $\boldsymbol\Psi_a$ is usually a positive diagonal uniqueness matrix, and $\phi_p(\cdot;\boldsymbol\mu,\boldsymbol\Sigma)$ denotes the multivariate normal density. Component $a$ admits the factor representation $\boldsymbol y_i=\boldsymbol\eta_a+\boldsymbol\Lambda_a\boldsymbol z_i+\boldsymbol\varepsilon_{ia}$, where $\boldsymbol z_i\sim\N_q(\boldsymbol0,\boldsymbol I)$ and $\boldsymbol\varepsilon_{ia}\sim\N_p(\boldsymbol0,\boldsymbol\Psi_a)$. This representation permits tractable conditional calculations even when the marginal density is multimodal.

A DGMM extends this representation to $h$ layers. Set $\boldsymbol z_i^{(0)}=\boldsymbol y_i$ and $r_0=p>r_1>\cdots>r_h$. The pathway $\boldsymbol S_i=(S_{i1},\ldots,S_{ih})$ records the component at each layer, with $S_{i\ell}\in\{1,\ldots,k_\ell\}$. Under independent layer allocations, $\pi_{\boldsymbol s}=\prod_{\ell=1}^{h}\pi_{s_\ell}^{(\ell)}$. Conditional on a pathway,
\begin{equation}\label{eq:transitions}
 \boldsymbol z_i^{(\ell-1)}=\boldsymbol\eta_{s_\ell}^{(\ell)}+
 \boldsymbol\Lambda_{s_\ell}^{(\ell)}\boldsymbol z_i^{(\ell)}+
 \boldsymbol\varepsilon_{i,s_\ell}^{(\ell)},\quad
 \boldsymbol\varepsilon_{i,s_\ell}^{(\ell)}\sim
 \N_{r_{\ell-1}}\bigl(\boldsymbol0,\boldsymbol\Psi_{s_\ell}^{(\ell)}\bigr),
\end{equation}
where disturbances are independent and $\boldsymbol z_i^{(h)}\sim\N_{r_h}(\boldsymbol0,\boldsymbol I)$. Define $\boldsymbol A_{\boldsymbol s}^{(0)}=\boldsymbol I_p$ and $\boldsymbol A_{\boldsymbol s}^{(\ell)}=\boldsymbol\Lambda_{s_1}^{(1)}\cdots\boldsymbol\Lambda_{s_\ell}^{(\ell)}$. Integrating through the hierarchy gives the exact pathway moments
\begin{equation}\label{eq:collapsed}
\begin{aligned}
 \boldsymbol\mu_{\boldsymbol s}&=\sum_{\ell=1}^{h}
 \boldsymbol A_{\boldsymbol s}^{(\ell-1)}\boldsymbol\eta_{s_\ell}^{(\ell)},\\
 \boldsymbol\Sigma_{\boldsymbol s}&=\boldsymbol A_{\boldsymbol s}^{(h)}\boldsymbol A_{\boldsymbol s}^{(h)\trans}+
 \sum_{\ell=1}^{h}\boldsymbol A_{\boldsymbol s}^{(\ell-1)}
 \boldsymbol\Psi_{s_\ell}^{(\ell)}\boldsymbol A_{\boldsymbol s}^{(\ell-1)\trans}.
\end{aligned}
\end{equation}
The induced observed density is
\begin{equation}\label{eq:dgmm_density}
 f_{\mathrm{DGMM}}(\boldsymbol y)=\sum_{\boldsymbol s\in\mathcal S}
 \pi_{\boldsymbol s}\phi_p(\boldsymbol y;\boldsymbol\mu_{\boldsymbol s},\boldsymbol\Sigma_{\boldsymbol s}),
 \qquad |\mathcal S|=K=\prod_{\ell=1}^{h}k_\ell.
\end{equation}
Although the hierarchy induces $K$ pathway densities, it shares their transition parameters. Marginalising the posterior pathway probabilities yields cluster memberships at each layer.

\subsection{Robust and asymmetric continuous pathways}\label{sec:continuous}

Normal scale mixtures connect Gaussian and heavy-tailed pathway distributions~\citep{AndrewsMallows1974, PeelMcLachlan2000}. Let $U_i=W_i^{-1}$ denote a shared precision. In an RDMM, conditional on $\boldsymbol S_i=\boldsymbol s$,
\begin{equation}\label{eq:rdmm}
 U_i\sim\Ga(\nu_{\boldsymbol s}/2,\nu_{\boldsymbol s}/2),\qquad
 \boldsymbol y_i\mid (U_i=u,\boldsymbol s)\sim
 \N_p\bigl(\boldsymbol\mu_{\boldsymbol s},\boldsymbol\Sigma_{\boldsymbol s}/u\bigr).
\end{equation}
The gamma distribution is parameterised by shape and rate. A common $u$ scales the deepest factor and the disturbance covariances throughout the hierarchy, yielding the marginal Student-$t$ density $f(\boldsymbol y_i\mid\boldsymbol s)=t_p(\boldsymbol y_i;\boldsymbol\mu_{\boldsymbol s},\boldsymbol\Sigma_{\boldsymbol s},\nu_{\boldsymbol s})$, where $\boldsymbol\Sigma_{\boldsymbol s}$ is a scale matrix. The shared precision preserves the Student-$t$ form under marginalisation, whereas independent layer-specific scales generally lead to a different marginal distribution. The Gaussian case is recovered as $\nu_{\boldsymbol s}\to\infty$~\citep{WuMcLachlan2026Robust}.

DStMMs additionally represent asymmetry through the normal mean--variance construction
\begin{equation}\label{eq:dstmm}
 \boldsymbol y_i\mid (U_i=u,\boldsymbol s)\sim\N_p\left(
 \boldsymbol\mu_{\boldsymbol s}+\frac{\boldsymbol\alpha_{\boldsymbol s}}{u},
 \frac{\boldsymbol\Sigma_{\boldsymbol s}}{u}\right),\quad
 \boldsymbol\alpha_{\boldsymbol s}=
 \sum_{\ell=1}^{h}\boldsymbol A_{\boldsymbol s}^{(\ell-1)}
 \boldsymbol\delta_{s_\ell}^{(\ell)}.
\end{equation}
Here $\boldsymbol\delta_{s_\ell}^{(\ell)}$ denotes the displacement coefficient of transition $\ell$, corresponding to a conditional shift $\boldsymbol\delta_{s_\ell}^{(\ell)}/u$. With the gamma precision in Eq.~\eqref{eq:rdmm}, this construction yields a generalised-hyperbolic skew-$t$ pathway~\citep{WuWangMcLachlan2026SkewT}, one of several multivariate skew-$t$ parameterisations~\citep{AzzaliniCapitanio2003, LeeMcLachlan2014}. Setting $\boldsymbol\alpha_{\boldsymbol s}=\boldsymbol0$ recovers the symmetric Student-$t$ case; for $\nu_{\boldsymbol s}>2$, the pathway mean is $\boldsymbol\mu_{\boldsymbol s}+\nu_{\boldsymbol s}\boldsymbol\alpha_{\boldsymbol s}/(\nu_{\boldsymbol s}-2)$. Related computational methods arise in generalised hyperbolic mixtures~\citep{BrowneMcNicholas2015}.

\subsection{Mixed-response deep mixtures}\label{sec:mixed}

For mixed observations, partition $\boldsymbol y_i=(\boldsymbol y_i^C,\boldsymbol y_i^D)$ into continuous and discrete responses. Their conditional measurement distributions are linked through continuous latent factors. A generic conditional-independence representation is
\begin{equation}\label{eq:mixedlink}
 p(\boldsymbol y_i\mid\boldsymbol z_i^C,\boldsymbol z_i^D)=
 \prod_{j\in\mathcal C}f_j(y_{ij}\mid\boldsymbol z_i^C)
 \prod_{j\in\mathcal D}g_j(y_{ij}\mid\boldsymbol z_i^D),
\end{equation}
where Gaussian distributions may be used for continuous variables and $g_j$ is chosen according to the discrete response type~\citep{McCullaghNelder1989, MoustakiKnott2000}. Typical choices include Bernoulli-logit models for binary responses, Poisson-log models for counts, threshold models for ordinal responses, and multinomial logits for nominal responses. These links allow different measurement scales to enter a joint latent-variable model.

MDGMMs use a common latent head (M1DGMM) or separate continuous and discrete heads connected by a shared deeper tail (M2DGMM)~\citep{FuchsPommeretViroli2022}. In either case, the observed likelihood can be written schematically as
\begin{equation}\label{eq:mixedlik}
 p(\boldsymbol y_i)=\sum_{\boldsymbol s}\pi_{\boldsymbol s}
 \int p(\boldsymbol y_i\mid\boldsymbol z,\boldsymbol s)
 p(\boldsymbol z\mid\boldsymbol s)\,d\boldsymbol z.
\end{equation}
The architecture determines the latent dimensions and the dependence between the two measurement blocks. Discrete likelihood terms generally retain a latent integral, in contrast to the closed-form pathway density for continuous Gaussian observations.

\section{Unified framework and likelihood-based inference}\label{sec:computing}

\subsection{Model dimensions and statistical specification}\label{sec:unified}

The specifications can be organised along three dimensions:
\begin{equation*}
\begin{aligned}
D&\in\{\mathrm{continuous},\mathrm{mixed}\},\\
F&\in\{\mathrm{Gaussian},t,\mathrm{skew}\mbox{-}t\},\\
A&\in\{\mathrm{single\ head},\mathrm{two\ head}\}.
\end{aligned}
\end{equation*}
The architecture dimension applies to mixed-response specifications: a single-head model uses a common latent representation, whereas a two-head model connects continuous and discrete representations through a shared tail. For continuous-only observations, the implementation uses the single-head structure.

Let $\boldsymbol S_i=\boldsymbol s$ denote the mixture pathway, $\boldsymbol Z_i$ the collection of continuous latent factors, and $U_i$ an optional precision variable. Conditional on $\boldsymbol Z_i=\boldsymbol z$, write $\boldsymbol z_C$ and $\boldsymbol z_D$ for the latent coordinates associated with the two response blocks. For continuous-only data, the discrete block $\boldsymbol y_i^D$ is empty. Denote the family-specific precision distribution by $g_F(u\mid\boldsymbol s)$ and the latent hierarchy by $h_F(\boldsymbol z\mid u,\boldsymbol s)$. For Gaussian kernels, $U_i=1$; Student-$t$ kernels use a gamma precision, and the skew-$t$ construction additionally introduces a displacement proportional to $U_i^{-1}$. The joint model then factorises as
\begin{equation}\label{eq:joint_framework}
 p(\boldsymbol y_i,\boldsymbol z,u,\boldsymbol s;\boldsymbol\theta)
 =\pi_{\boldsymbol s}\,g_F(u\mid\boldsymbol s)\,
 h_F(\boldsymbol z\mid u,\boldsymbol s)\,
 f_C(\boldsymbol y_i^C\mid\boldsymbol z_C,u,\boldsymbol s)\,
 f_D(\boldsymbol y_i^D\mid\boldsymbol z_D,\boldsymbol s).
\end{equation}
Here $f_C$ is the conditional Gaussian measurement density and $f_D$ is the product of response-specific discrete measurement probabilities. In a single-head model, $\boldsymbol z_C$ and $\boldsymbol z_D$ coincide; in a two-head model, they are linked through a deeper shared factor. Marginalising the unobserved variables gives
\begin{equation}\label{eq:observed_framework}
\begin{aligned}
 L(\boldsymbol\theta)=\prod_{i=1}^n\Bigg\{
 &\sum_{\boldsymbol s\in\mathcal S}\pi_{\boldsymbol s}
 \int\!\int f_C(\boldsymbol y_i^C\mid\boldsymbol z_C,u,\boldsymbol s) \\
 &\quad\times f_D(\boldsymbol y_i^D\mid\boldsymbol z_D,\boldsymbol s)
 h_F(\boldsymbol z\mid u,\boldsymbol s)\,g_F(u\mid\boldsymbol s)
 \,d\boldsymbol z\,du\Bigg\}.
\end{aligned}
\end{equation}
For Gaussian kernels, integration over $u$ is replaced by evaluation at $u=1$. With continuous measurements, marginalisation produces the pathway densities in Section~\ref{sec:models}; with discrete measurements, the response likelihood remains inside the latent integral and typically requires numerical approximation.

For discrete variable $j$, let $\eta_{ij}=\beta_{0j}+\boldsymbol\beta_j\trans\boldsymbol z_D$ be a conditional linear predictor. Common links include
\begin{equation}\label{eq:links_framework}
 \Pr(Y_{ij}=1\mid\boldsymbol z_D)=\operatorname{logit}^{-1}(\eta_{ij}),\qquad
 \E(Y_{ij}\mid\boldsymbol z_D)=\exp(\eta_{ij})\quad\text{for counts}.
\end{equation}
Ordinal variables may use ordered cumulative-link thresholds, whereas nominal variables can use multinomial logits relative to a reference category. Identification constraints depend on the response family.

\begin{table}[!ht]
\caption{Model families and mixed-response specifications in the unified framework.}\label{tab:framework}
\centering\small\renewcommand{\arraystretch}{1.18}
\begin{tabularx}{\textwidth}{@{}p{23mm}p{22mm}p{27mm}X@{}}
\toprule
Response & Kernel & Architecture & Model family \\
\midrule
Continuous & Gaussian & Single-head & DGMM \citep{ViroliMcLachlan2019} \\
Continuous & Student-$t$ & Single-head & RDMM \citep{WuMcLachlan2026Robust} \\
Continuous & Skew-$t$ & Single-head & DStMM \citep{WuWangMcLachlan2026SkewT} \\
Mixed & Gaussian & Single- or two-head & MDGMM \citep{FuchsPommeretViroli2022} \\
Mixed & Student-$t$ & Single- or two-head & Robust mixed-response  \\
Mixed & Skew-$t$ & Single- or two-head & Asymmetric mixed-response\\
\bottomrule
\end{tabularx}
\end{table}

Table~\ref{tab:framework} cross-classifies the reviewed families and the corresponding mixed-response specifications. In the mixed Student-$t$ and skew-$t$ settings, response-specific measurement links are combined with heavy-tailed or asymmetric latent kernels. The response links and the latent kernel thus define distinct parts of the joint distribution.

\subsection{Estimation strategy and the EM objective}\label{sec:estimate_unified}

The families share an expectation--maximisation (EM) framework~\citep{mclachlan2008algorithm} but differ in the conditional moments required by the E-step. Let $\boldsymbol H_i=(\boldsymbol S_i,\boldsymbol Z_i,U_i)$ collect the missing variables, with $U_i$ omitted in the Gaussian case. For continuous responses, Gaussian, Student-$t$, and skew-$t$ kernels lead to Gaussian, gamma, and generalised inverse Gaussian (GIG) calculations, respectively. Mixed-response models generally require numerical approximation, such as Monte Carlo EM (MCEM)~\citep{WeiTanner1990}, and response-specific optimisation in the M-step. The pathway posterior provides a common starting point for the E-step across these specifications.

\subsection{Observed likelihood and pathway posterior probabilities}\label{sec:posterior}

Let $q_{\boldsymbol s}(\boldsymbol y_i;\boldsymbol\theta)$ denote the observed density or probability mass within pathway $\boldsymbol s$, with the appropriate product measure for mixed observations. It is available in closed form for the continuous Gaussian, Student-$t$, and generalised-hyperbolic skew-$t$ models, and is obtained from the latent integral in Eq.~\eqref{eq:observed_framework} for mixed responses. The log likelihood and posterior pathway probabilities are
\begin{equation}\label{eq:ll_tau}
 \ell(\boldsymbol\theta)=\sum_{i=1}^n\log\left\{\sum_{\boldsymbol s\in\mathcal S}
 \pi_{\boldsymbol s}q_{\boldsymbol s}(\boldsymbol y_i;\boldsymbol\theta)\right\},\qquad
 \tau_{i\boldsymbol s}=
 \frac{\pi_{\boldsymbol s}q_{\boldsymbol s}(\boldsymbol y_i;\boldsymbol\theta)}
 {\sum_{\boldsymbol t\in\mathcal S}\pi_{\boldsymbol t}q_{\boldsymbol t}(\boldsymbol y_i;\boldsymbol\theta)}.
\end{equation}
For numerical stability, set $a_{i\boldsymbol s}=\log\pi_{\boldsymbol s}+\log q_{\boldsymbol s}(\boldsymbol y_i)$ and evaluate $\log\sum_{\boldsymbol s}\exp(a_{i\boldsymbol s})=m_i+\log\sum_{\boldsymbol s}\exp(a_{i\boldsymbol s}-m_i)$, where $m_i=\max_{\boldsymbol s}a_{i\boldsymbol s}$. Under independent layer allocations, marginalising the pathway responsibilities gives layer-specific weights $\tau_{i,\ell,a}=\sum_{\boldsymbol s:s_\ell=a}\tau_{i\boldsymbol s}$ and the mixing-proportion update
\begin{equation}\label{eq:piupdate}
 \widehat\pi_a^{(\ell)}=\frac{1}{n}\sum_{i=1}^n\tau_{i,\ell,a}.
\end{equation}
For the complete latent-variable model, the EM objective is
\begin{equation}\label{eq:qfunc}
 Q(\boldsymbol\theta\mid\boldsymbol\theta^{(t)})=
 \sum_{i=1}^n\E_{\boldsymbol\theta^{(t)}}\left[
 \log p(\boldsymbol y_i,\boldsymbol H_i;\boldsymbol\theta)
 \mid\boldsymbol y_i\right].
\end{equation}
The expectations in Eq.~\eqref{eq:qfunc} depend on the pathway kernel and observation model~\citep{DempsterLairdRubin1977}.

\subsection{Gaussian conditional moments and factor updates}\label{sec:gaussiancomp}

For a fixed DGMM pathway, stack the latent factors as $\boldsymbol Z_i=(\boldsymbol z_i^{(1)\trans},\ldots,\boldsymbol z_i^{(h)\trans})\trans$. Since $(\boldsymbol y_i\trans,\boldsymbol Z_i\trans)\trans$ is jointly normal, the pathway-specific moments $\boldsymbol m_{Z,\boldsymbol s}$, $\boldsymbol C_{Zy,\boldsymbol s}$, and $\boldsymbol C_{ZZ,\boldsymbol s}$ give
\begin{equation}\label{eq:conditional_g}
\begin{aligned}
 \boldsymbol m_{i\boldsymbol s}&=\E(\boldsymbol Z_i\mid\boldsymbol y_i,\boldsymbol s)
 =\boldsymbol m_{Z,\boldsymbol s}+\boldsymbol C_{Zy,\boldsymbol s}
 \boldsymbol\Sigma_{\boldsymbol s}^{-1}(\boldsymbol y_i-\boldsymbol\mu_{\boldsymbol s}),\\
 \boldsymbol V_{\boldsymbol s}&=\Var(\boldsymbol Z_i\mid\boldsymbol y_i,\boldsymbol s)
 =\boldsymbol C_{ZZ,\boldsymbol s}-\boldsymbol C_{Zy,\boldsymbol s}
 \boldsymbol\Sigma_{\boldsymbol s}^{-1}\boldsymbol C_{yZ,\boldsymbol s}.
\end{aligned}
\end{equation}
The corresponding second moment is $\E(\boldsymbol Z_i\boldsymbol Z_i\trans\mid\boldsymbol y_i,\boldsymbol s)=\boldsymbol V_{\boldsymbol s}+\boldsymbol m_{i\boldsymbol s}\boldsymbol m_{i\boldsymbol s}\trans$. Posterior cross-moments between adjacent layers provide the additional sufficient statistics for the shared factor transitions.

For layer $\ell$ and component $a$, define $\boldsymbol x_i^{(\ell)}=(1,\boldsymbol z_i^{(\ell)\trans})\trans$ and $\boldsymbol B_a^{(\ell)}=[\boldsymbol\eta_a^{(\ell)},\boldsymbol\Lambda_a^{(\ell)}]$. Let $\E_{i\boldsymbol s}(\cdot)$ denote expectation conditional on $(\boldsymbol y_i,\boldsymbol S_i=\boldsymbol s)$ at the current parameter values. Weighted least squares yields
\begin{equation}\label{eq:Bupdate}
 \widehat{\boldsymbol B}_a^{(\ell)}=
 \left\{\sum_{i,\boldsymbol s:s_\ell=a}\tau_{i\boldsymbol s}
 \E_{i\boldsymbol s}(\boldsymbol z_i^{(\ell-1)}\boldsymbol x_i^{(\ell)\trans})\right\}
 \left\{\sum_{i,\boldsymbol s:s_\ell=a}\tau_{i\boldsymbol s}
 \E_{i\boldsymbol s}(\boldsymbol x_i^{(\ell)}\boldsymbol x_i^{(\ell)\trans})\right\}^{-1}.
\end{equation}
The diagonal uniqueness update is the diagonal of the weighted expected residual cross-product, divided by $\sum_{i,\boldsymbol s:s_\ell=a}\tau_{i\boldsymbol s}$. Loading or uniqueness restrictions require constrained maximisation. Gaussian conditioning or smoothing provides the necessary moments, although ill-conditioned factor blocks may require regularisation.

\subsection{Shared-scale inference for robust pathways}\label{sec:tcomp}

Let $d_{i\boldsymbol s}=(\boldsymbol y_i-\boldsymbol\mu_{\boldsymbol s})\trans\boldsymbol\Sigma_{\boldsymbol s}^{-1}(\boldsymbol y_i-\boldsymbol\mu_{\boldsymbol s})$ be the squared Mahalanobis distance. The normal--gamma representation in Eq.~\eqref{eq:rdmm} gives
\begin{equation}\label{eq:tposterior}
 U_i\mid(\boldsymbol y_i,\boldsymbol s)\sim
 \Ga\left(\frac{\nu_{\boldsymbol s}+p}{2},\frac{\nu_{\boldsymbol s}+d_{i\boldsymbol s}}{2}\right),\quad
 \E(U_i\mid\boldsymbol y_i,\boldsymbol s)=\frac{\nu_{\boldsymbol s}+p}{\nu_{\boldsymbol s}+d_{i\boldsymbol s}}.
\end{equation}
The corresponding log-precision moment is
\begin{equation}\label{eq:elogU}
 \E(\log U_i\mid\boldsymbol y_i,\boldsymbol s)=
 \psi\left(\frac{\nu_{\boldsymbol s}+p}{2}\right)
 -\log\left(\frac{\nu_{\boldsymbol s}+d_{i\boldsymbol s}}{2}\right).
\end{equation}
Here $\psi$ denotes the digamma function. For the factor-transition updates, the Gaussian sufficient statistics in Eq.~\eqref{eq:Bupdate} become $\E_{i\boldsymbol s}(U_i\boldsymbol z_i^{(\ell-1)}\boldsymbol x_i^{(\ell)\trans})$ and $\E_{i\boldsymbol s}(U_i\boldsymbol x_i^{(\ell)}\boldsymbol x_i^{(\ell)\trans})$. The weighted normal-equation structure is retained, but its moments are evaluated under the joint latent posterior.

Maximising the expected gamma log density gives the following stationarity equation for a degrees-of-freedom parameter $\nu$ shared across a specified set of pathways:
\begin{equation}\label{eq:nuupdate}
 \log(\nu/2)-\psi(\nu/2)+1+
 \frac{\sum_{i,\boldsymbol s}\tau_{i\boldsymbol s}
 \{\E_{i\boldsymbol s}(\log U_i)-\E_{i\boldsymbol s}(U_i)\}}
 {\sum_{i,\boldsymbol s}\tau_{i\boldsymbol s}}=0.
\end{equation}
The sums in Eq.~\eqref{eq:nuupdate} cover the pathways sharing $\nu$. Bounded scalar optimisation is useful when effective component sizes are small. Observations with large Mahalanobis distances receive lower expected precisions, which underlies the robust-mixture behaviour~\citep{LangeLittleTaylor1989, PeelMcLachlan2000}. Estimation of the full RDMM hierarchy also involves stochastic or Monte Carlo steps~\citep{WuMcLachlan2026Robust}. The \texttt{MixedDeepMM} implementation uses a common $\nu$ across pathways, reducing the number of degrees-of-freedom parameters.

\subsection{Generalised inverse Gaussian computations for skewed pathways}\label{sec:skewcomp}

The skew-$t$ construction in Eq.~\eqref{eq:dstmm} is conditionally Gaussian, but the posterior precision is no longer gamma. Define
\begin{equation}\label{eq:gigpar}
 \chi_{\boldsymbol s}=\boldsymbol\alpha_{\boldsymbol s}\trans\boldsymbol\Sigma_{\boldsymbol s}^{-1}\boldsymbol\alpha_{\boldsymbol s},\qquad
 \omega_{i\boldsymbol s}=\nu_{\boldsymbol s}+d_{i\boldsymbol s},\qquad
 \lambda_{\boldsymbol s}=\frac{\nu_{\boldsymbol s}+p}{2}.
\end{equation}
Combining the conditional Gaussian likelihood with the gamma prior gives
\begin{equation}\label{eq:gigposterior}
 p(u\mid\boldsymbol y_i,\boldsymbol s)\propto
 u^{\lambda_{\boldsymbol s}-1}
 \exp\left\{-\frac{1}{2}\left(\omega_{i\boldsymbol s}u+
 \frac{\chi_{\boldsymbol s}}{u}\right)\right\},\qquad u>0.
\end{equation}
Thus $U_i\mid(\boldsymbol y_i,\boldsymbol s)\sim\GIG(\lambda_{\boldsymbol s},\chi_{\boldsymbol s},\omega_{i\boldsymbol s})$, with the parameterisation defined by Eq.~\eqref{eq:gigposterior}. Its conditional power moments are
\begin{equation}\label{eq:gigmoments}
 \E(U_i^r\mid\boldsymbol y_i,\boldsymbol s)=
 \left(\frac{\chi_{\boldsymbol s}}{\omega_{i\boldsymbol s}}\right)^{r/2}
 \frac{K_{\lambda_{\boldsymbol s}+r}\bigl(\sqrt{\chi_{\boldsymbol s}\omega_{i\boldsymbol s}}\bigr)}
 {K_{\lambda_{\boldsymbol s}}\bigl(\sqrt{\chi_{\boldsymbol s}\omega_{i\boldsymbol s}}\bigr)},
 \qquad \chi_{\boldsymbol s}>0.
\end{equation}
Here $K_\lambda$ denotes the modified Bessel function of the second kind. The moments for $r=1$ and $r=-1$, together with $\E(\log U_i\mid\boldsymbol y_i,\boldsymbol s)$, enter parameter estimation. At $\chi_{\boldsymbol s}=0$, the posterior and its moments follow directly from the gamma limit. Log-scale evaluation or specialised GIG routines improve stability~\citep{BrowneMcNicholas2015, WuWangMcLachlan2026SkewT}.

The $U_i^{-1}$ displacement introduces additional scale-weighted cross-moments in the transition updates. Distinct layer-wise displacements can also induce the same collapsed $\boldsymbol\alpha_{\boldsymbol s}$, linking the estimation of asymmetry to the identifiability of the latent hierarchy.

\subsection{Latent integration for mixed measurements}\label{sec:mixedcomp}

For mixed responses, write $\boldsymbol v=\boldsymbol z$ for Gaussian pathways and $\boldsymbol v=(\boldsymbol z,u)$ for Student-$t$ and skew-$t$ pathways. The latent posterior within a pathway follows from Eq.~\eqref{eq:joint_framework}:
\begin{equation}\label{eq:mixedposterior}
 p(\boldsymbol v\mid\boldsymbol y_i,\boldsymbol s)
 \propto p(\boldsymbol y_i\mid\boldsymbol v,\boldsymbol s)
 p(\boldsymbol v\mid\boldsymbol s).
\end{equation}
The discrete measurement terms generally preclude Gaussian conditioning. Numerical alternatives include importance sampling, posterior simulation, and Laplace approximation. For draws $\boldsymbol v_{im}\sim q_{i\boldsymbol s}(\boldsymbol v)$, the self-normalised importance approximation to a posterior expectation is
\begin{equation}\label{eq:IS}
\begin{aligned}
 \widehat{\E}\{h(\boldsymbol V_i)\mid\boldsymbol y_i,\boldsymbol s\}
 &=\frac{\sum_{m=1}^M w_{im\boldsymbol s}h(\boldsymbol v_{im})}
 {\sum_{m=1}^M w_{im\boldsymbol s}},\\
 w_{im\boldsymbol s}&=
 \frac{p(\boldsymbol y_i\mid\boldsymbol v_{im},\boldsymbol s)
 p(\boldsymbol v_{im}\mid\boldsymbol s)}{q_{i\boldsymbol s}(\boldsymbol v_{im})}.
\end{aligned}
\end{equation}
The same latent integration determines the pathway density in Eq.~\eqref{eq:ll_tau}. MCEM and related stochastic methods approximate the conditional expectations in the EM objective~\citep{WeiTanner1990,Nielsen2000,FuchsPommeretViroli2022}. For mixed-response estimation, \texttt{MixedDeepMM} uses MCEM with structured data-derived initialisation~\citep{MixedDeepMM2026}.

\subsection{Clustering assessment and model-selection considerations}\label{sec:selection}

Clustering in a deep mixture is defined at a particular level of the hierarchy: a complete pathway, an individual mixture layer, or the shared tail of a two-head architecture. At a given level, the posterior probabilities determine a maximum-posterior-probability partition. External agreement is quantified by the adjusted Rand index (ARI)~\citep{HubertArabie1985} and the misclassification rate (MR) after optimal label matching. These measures use external labels retained for evaluation after fitting.

For mixed observations, we additionally use the average silhouette width~\citep{Rousseeuw1987} based on Gower dissimilarity~\citep{Gower1971}. This dissimilarity accommodates variable-specific measurement types: continuous and count variables contribute range-scaled numerical differences, binary and nominal variables contribute categorical mismatches, and ordinal variables retain their ordering. Given a partition and the predictor-based Gower dissimilarities, let $a(i)$ denote the average dissimilarity of observation $i$ to its assigned cluster, and let $b(i)$ denote the smallest average dissimilarity to another cluster. Its silhouette width is
\begin{equation}\label{eq:gower_silhouette}
 s(i)=\frac{b(i)-a(i)}{\max\{a(i),b(i)\}} \qquad \text{and} \qquad 
 \overline{s}=\frac{1}{n}\sum_{i=1}^{n}s(i).
\end{equation}
For a singleton cluster, $s(i)=0$; the average silhouette is defined for partitions with at least two occupied clusters (and fewer clusters than observations). Larger values indicate greater within-cluster cohesion relative to between-cluster separation under the chosen dissimilarity. In the Cryotherapy application, the same predictor-based Gower dissimilarity is used for all partitions.

Likelihood-based criteria describe a complementary aspect of model fit. The Bayesian information criterion is $\mathrm{BIC}=-2\widehat\ell+d\log n$, where $d$ denotes the number of estimated parameters~\citep{Schwarz1978}. In a deep mixture, parameter sharing and identifiability constraints determine the parameter count, while numerical integration in mixed-response models yields an approximate BIC. The applications use BIC to summarise likelihood-based fit alongside partition-based measures. The joint selection of component numbers, factor dimensions, and head architecture remains an open problem.

\section{Illustrative analyses of two real datasets}\label{sec:applications}

Two publicly available UCI datasets illustrate the framework: Wine, with continuous chemical measurements, and Cryotherapy, with continuous, binary, nominal, and count predictors. The class labels and treatment outcomes are retained for external evaluation, and the models are fitted to the predictors alone. The analyses report ARI, MR, and BIC for both datasets, together with the Gower silhouette for Cryotherapy.

For each configuration, $R=20$ fits use seeds 42--61 and one initialisation per seed. The analysis scripts specify a maximum of 100 iterations, a minimum of eight iterations, a tolerance of 0.002, an initial degrees-of-freedom value of 8 for non-Gaussian kernels, and a uniqueness floor of 0.03. The mixed-response configurations additionally specify 48 Monte Carlo samples for estimation, 192 for likelihood evaluation, and a ridge parameter of 0.001; the continuous skew-$t$ configuration specifies 48 Monte Carlo samples.

For a metric $m$, the run mean is $\overline{m}=R^{-1}\sum_{r=1}^{R}m_r$ and its standard error is $\operatorname{SE}(\overline{m})=\operatorname{sd}(m_1,\ldots,m_R)/\sqrt{R}$. These standard errors quantify the precision of the run mean over repeated fits to a fixed dataset. Each model's consensus partition is constructed from the co-assignment frequencies
\begin{equation}\label{eq:coassignment}
 C_{ij}=\frac{1}{R}\sum_{r=1}^{R}
 \mathbb{I}\bigl\{g_i^{(r)}=g_j^{(r)}\bigr\},
\end{equation}
where $g_i^{(r)}$ is the cluster label of observation $i$ in run $r$. Average-linkage hierarchical clustering of $1-C_{ij}$ gives a consensus partition at the target number of occupied clusters (three for Wine and two for Cryotherapy in these fits). This construction is invariant to permutations of the cluster labels across runs. The figures display the consensus partitions, and the tables summarise the individual runs.

\subsection{Continuous measurements: UCI Wine}\label{sec:wine}

The UCI Wine dataset contains $178$ observations from three wine classes, measured on $13$ continuous chemical variables. All predictors are standardised to zero mean and unit sample standard deviation. The wine-class labels are excluded from estimation and are used only to assess the fitted clusters. We consider the Gaussian DGMM, Student-$t$ RDMM, and skew-$t$ DStMM with a two-layer, single-head structure, component counts $\boldsymbol{k}=(3,2)$, and latent dimensions $\boldsymbol{r}=(4,2)$. For the Student-$t$ and skew-$t$ fits, a common degrees-of-freedom parameter $\nu$ is estimated across pathways. Clustering is evaluated at the first mixture layer using maximum-posterior-probability assignments. All runs produced three occupied clusters.

\begin{table}[!htbp]
\caption{UCI Wine clustering results (mean $\pm$ SE). }
\label{tab:wine_results}
\centering\small
\setlength{\tabcolsep}{5pt}
\renewcommand{\arraystretch}{1.15}
\begin{tabular}{@{}lcccc@{}}
\toprule
Model & ARI & MR & BIC & $\hat{\nu}$ \\
\midrule
DGMM  & $0.9149\pm0.0000$ & $0.0281\pm0.0000$ & $5652.31\pm0.78$ & --- \\
RDMM  & $0.9832\pm0.0000$ & $0.0056\pm0.0000$ & $5538.40\pm0.00$ & $20.81\pm0.00$ \\
DStMM & $0.9583\pm0.0041$ & $0.0140\pm0.0014$ & $5562.77\pm1.80$ & $19.51\pm0.14$ \\
\bottomrule
\end{tabular}
\end{table}

\begin{figure}[!htbp]
\centering
\includegraphics[width=\textwidth]{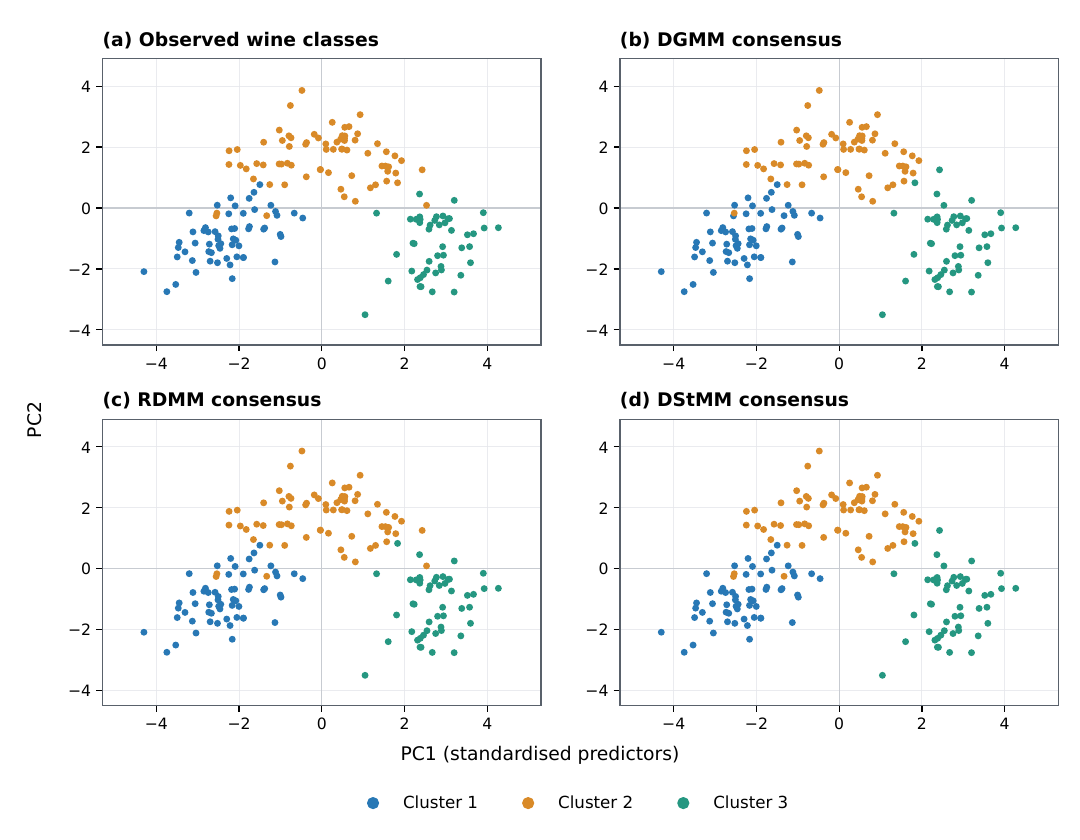}
\caption{Principal component projection of the standardised UCI Wine predictors. Panel (A) shows the observed wine classes; panels (B)--(D) show the DGMM, RDMM, and DStMM consensus partitions constructed from the fitted partitions. All panels use identical PCA coordinates and scales. Cluster identifiers and colours are specific to each partition.}
\label{fig:wine_results}
\end{figure}

In Table~\ref{tab:wine_results}, RDMM attains the highest mean ARI (0.9832), the lowest mean MR (0.0056), and the lowest mean BIC among the three specifications. DGMM and RDMM yield identical ARI and MR values across repeated fits to the reported precision, whereas DStMM exhibits some run-to-run variation. Figure~\ref{fig:wine_results} displays the corresponding consensus partitions in a common principal component projection. With the latent architecture held fixed, the comparison illustrates how the pathway distribution relates to likelihood-based fit and agreement with the observed wine classes.

\subsection{Mixed measurements: UCI Cryotherapy}\label{sec:cryotherapy}

The UCI Cryotherapy dataset contains $90$ observations and a binary treatment-outcome label. Its six predictors include three continuous variables (age, treatment time, and affected area), one count variable (number of warts), one binary variable (sex), and one nominal variable (wart type). Continuous predictors are standardised; discrete predictors retain their respective measurement types. The treatment outcome provides an external clinical response; the Gower dissimilarity and model estimation use the six predictors.

We compare Gaussian, Student-$t$, and skew-$t$ kernels under single-head (Single) and two-head (Double) architectures, retaining the shorthand DGMM, RDMM, and DStMM for the respective kernel families. The single-head configurations use $\boldsymbol{k}=(2,2)$ and $\boldsymbol{r}=(2,1)$, with clustering evaluated at the first common mixture layer. The two-head configurations use $\boldsymbol{k}=(2,2,2)$, two-dimensional continuous and discrete heads, and a one-dimensional shared tail; clustering is evaluated at the shared-tail layer. All 20 runs for each configuration yielded two occupied clusters. Table~\ref{tab:cryotherapy} reports run-averaged Gower silhouette widths, ARI, MR, BIC, and estimated degrees of freedom.

\begin{table}[!ht]
\caption{UCI Cryotherapy clustering results (mean $\pm$ SE).}
\label{tab:cryotherapy}
\begin{tabular}{@{}lccccc@{}}
\toprule
Model & Gower silhouette & ARI & MR & BIC & $\hat{\nu}$ \\
\midrule
DGMM (Single)  & $0.1749\pm0.0122$ & $0.0463\pm0.0137$ & $0.4033\pm0.0153$ & $1484.17\pm5.04$ & --- \\
RDMM (Single)  & $0.1597\pm0.0029$ & $0.0560\pm0.0105$ & $0.3822\pm0.0118$ & $1545.19\pm0.96$ & $3.12\pm0.01$ \\
DStMM (Single) & $0.1595\pm0.0033$ & $0.0758\pm0.0134$ & $0.3639\pm0.0123$ & $1548.87\pm1.10$ & $3.37\pm0.02$ \\
\midrule
DGMM (Double)  & $0.2441\pm0.0038$ & $0.1191\pm0.0056$ & $0.3233\pm0.0038$ & $1508.85\pm1.67$ & --- \\
RDMM (Double)  & $0.2395\pm0.0039$ & $0.1129\pm0.0078$ & $0.3289\pm0.0057$ & $1509.59\pm1.79$ & $19.26\pm2.00$ \\
DStMM (Double) & $0.1967\pm0.0028$ & $0.2646\pm0.0193$ & $0.2428\pm0.0101$ & $1540.80\pm2.35$ & $5.43\pm0.14$ \\
\bottomrule
\end{tabular}%
\end{table}

\begin{figure}[!htbp]
\centering
\includegraphics[width=\textwidth]{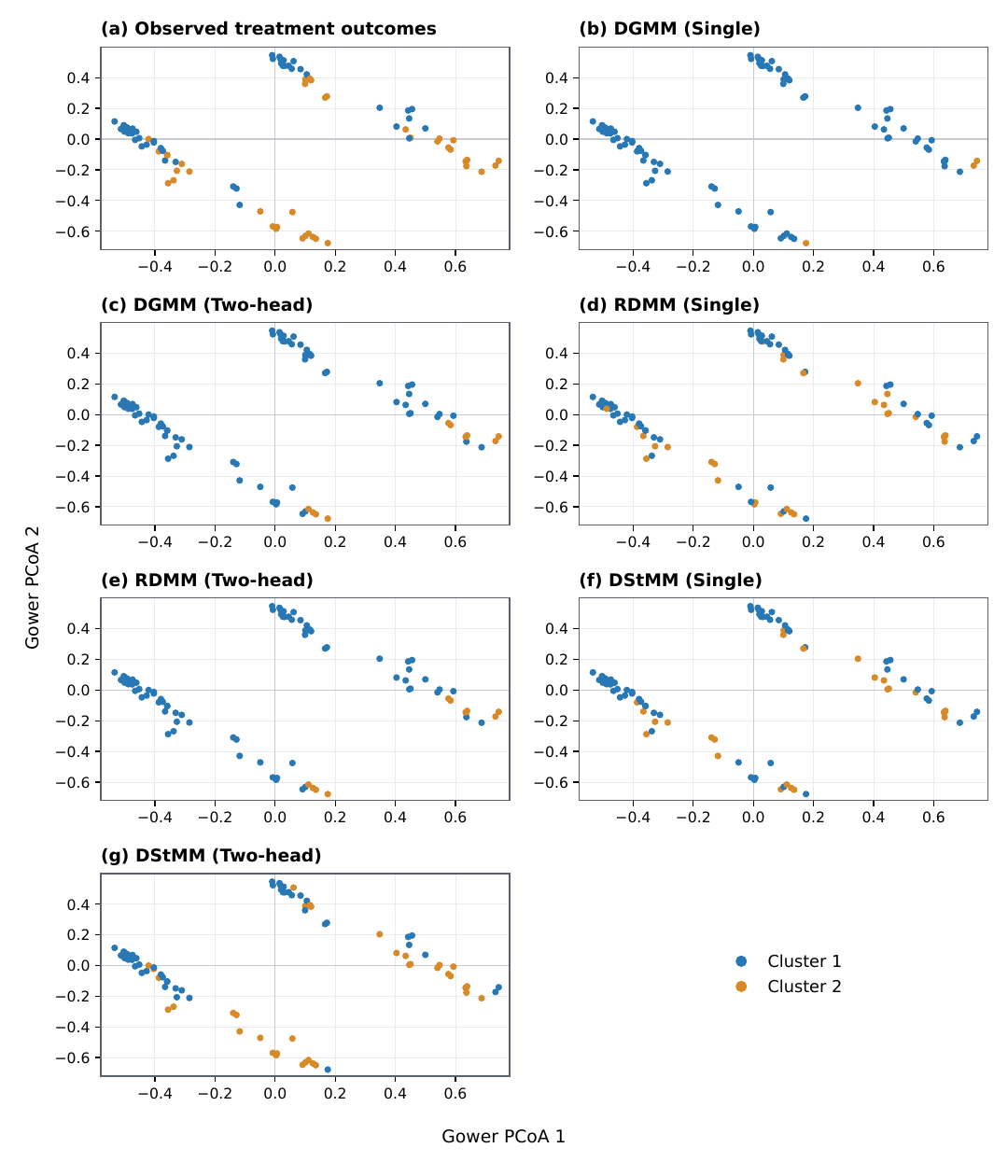}
\caption{Gower principal coordinates analysis (PCoA) of the Cryotherapy predictors, using an additive constant correction, with the observed treatment outcomes and six model-specific consensus partitions. All seven scatter panels share the same coordinates and axis ranges; the eighth panel contains the legend. Colours identify observed outcomes or predicted clusters within each panel.}
\label{fig:cryotherapy-gower-pcoa}
\end{figure}

In Table~\ref{tab:cryotherapy}, the two-head configurations yield higher mean Gower silhouette widths and ARI values than the corresponding single-head configurations for all three kernels. These comparisons involve both the head structure and the level at which the partition is extracted. DGMM (Double) has the highest mean silhouette (0.2441), whereas DStMM (Double) has the highest mean ARI (0.2646) and the lowest mean MR (0.2428). Internal separation in the predictor space and agreement with treatment outcomes thus capture different properties of the partitions.

Figure~\ref{fig:cryotherapy-gower-pcoa} further illustrates differences in cluster occupancy: the DGMM (Single) consensus contains $87$ and $3$ observations, compared with $51$ and $39$ for DStMM (Double). The Gaussian single-head specification has the lowest mean BIC, illustrating a further distinction between likelihood-based fit and partition characteristics. Together with the Wine analysis, these results connect the choice of pathway kernel and latent architecture to the different statistical objectives of model-based clustering.

\section{Discussion}\label{sec:discussion}

The hierarchical factor-mixture representation provides a common statistical basis for the model families reviewed here. Its defining feature is the sharing of transition parameters across latent pathways, with the pathway kernel governing tail behaviour and asymmetry. Gaussian hierarchies admit direct conditional calculations; the shared precision in Student-$t$ hierarchies introduces observation-specific weighting; and the skew-$t$ displacement adds further scale-dependent moments. In mixed-response models, these distributional features operate jointly with the measurement links, while the single- and two-head architectures determine how dependence across response types is represented. The resulting likelihood and posterior structure account for the different computational requirements of the model families. The two applications illustrate complementary aspects of this framework. Wine isolates the choice of pathway kernel within a fixed architecture, whereas Cryotherapy combines kernel and architecture variation in the presence of heterogeneous measurement types. The latter example also distinguishes internal cluster separation, agreement with an external clinical response, and likelihood-based fit. These quantities describe different statistical targets, while consensus partitions summarise the structure shared across repeated fits.

The number of possible pathways increases multiplicatively with the component counts across layers, even though the transition parameters are shared. This combination of flexibility and computational complexity connects architecture selection to parameter identifiability, component occupancy, and numerical stability. Variational inference and sparsity provide complementary approaches to high-dimensional deep mixtures~\citep{KockKleinNott2022}, while empty global components can occur even when all layer-specific components are occupied~\citep{SelosseEtAl2020}. For mixed-response models, numerical integration introduces a further source of variability in likelihood evaluation and posterior inference. Joint selection of component counts, latent dimensions, and head architecture, together with the assessment of partition stability and parameter uncertainty, constitutes a natural direction for further statistical research.

\appendix

\section{\texttt{MixedDeepMM} version 0.1.0: R interface}\label{app:r}

\begin{table}[htbp]
\caption{Principal functions in \texttt{MixedDeepMM}.}\label{tab:rfuns}
\centering\small
\renewcommand{\arraystretch}{1.12}
\begin{tabularx}{\textwidth}{@{}p{55mm}X@{}}
\toprule
Function & Purpose \\
\midrule
\texttt{mixeddeepmm()} & Fits a specified Gaussian, Student-$t$, or skew-$t$ hierarchy. \\
\texttt{summary()}, \texttt{logLik()} & Summarise model estimates and fitted likelihood. \\
\texttt{predict()} & Obtains layer-specific posterior cluster probabilities. \\
\texttt{mixeddeepmm\_ari()}, \texttt{mixeddeepmm\_mr()} & Evaluate agreement with external class labels. \\
\texttt{mixeddeepmm\_gower\_distance()}, \texttt{mixeddeepmm\_gower\_silhouette()} & Compute mixed-data dissimilarities and the corresponding silhouette. \\
\bottomrule
\end{tabularx}
\end{table}

The \texttt{MixedDeepMM} R package~\citep{MixedDeepMM2026} provides a common interface for Gaussian, Student-$t$, and skew-$t$ deep mixtures with continuous and mixed measurements. The principal functions are listed in Table~\ref{tab:rfuns}. Version 0.1.0 is available from CRAN at \url{https://cran.r-project.org/package=MixedDeepMM}.

\begin{lstlisting}[language=R]
install.packages("MixedDeepMM")
library(MixedDeepMM)
packageVersion("MixedDeepMM")
fit <- mixeddeepmm(
  data = X, types = types, family = "t", heads = "two",
  layers = list(k = c(2L, 2L, 2L),
                r = list(continuous = 2L,
                         discrete = 2L, tail = 1L)),
  control = list(seed = 42L, nstart = 1L,
                 max_iter = 100L, min_iter = 8L, tol = 0.002,
                 mc_samples = 48L, mc_eval = 192L,
                 nu = 8, estimate_nu = TRUE,
                 ridge = 0.001, psi_floor = 0.03)
)
post <- predict(fit, type = "layer",
                cluster_level = "paper")
clusters <- max.col(post, ties.method = "first")
\end{lstlisting}

The following example uses prepared predictors \texttt{X} and a named measurement-type vector \texttt{types} to specify the two-head Student-$t$ model in the Cryotherapy analysis. The installation command retrieves the current CRAN release; the version command identifies the installed package. For continuous measurements, \texttt{heads="single"} 
and \texttt{layers=list(k=c(3L,2L), r=c(4L,2L))} 
specify the Wine hierarchy, with the measurement-type 
vector omitted. The \texttt{family} argument selects 
the Gaussian, Student-$t$, or skew-$t$ pathway kernel. 
The \texttt{predict()} method returns posterior 
probabilities at the specified mixture layer; 
\texttt{cluster\_level="paper"} selects the shared 
tail in the two-head example.

\backmatter

\bmhead{Funding} This work was supported by the Australian Research Council [DP230101671].

\bmhead{Competing interests} The authors declare that they have no competing interests.

\bmhead{Data availability}
The Wine and Cryotherapy datasets are publicly available 
from the UCI Machine Learning Repository.

\bmhead{Code availability}
The \texttt{MixedDeepMM} R package (version 0.1.0) is 
available from CRAN at 
\url{https://cran.r-project.org/package=MixedDeepMM}.
The processed datasets, analysis scripts, and accompanying 
results are available at 
\url{https://github.com/wujrtudou/MixedDeepMMcases}.

\bmhead{Author contributions} Jinran Wu: Writing -- review \& editing, Writing -- original draft, Visualization, Software, Methodology, Formal analysis, Conceptualization. Geoffrey J. McLachlan: Supervision, Investigation, Conceptualization, Methodology, Writing -- review \& editing, Funding acquisition.

\bibliography{refs}

@article{DempsterLairdRubin1977,
 author={Dempster, Arthur P. and Laird, Nan M. and Rubin, Donald B.},
 title={Maximum likelihood from incomplete data via the {EM} algorithm},
 journal={Journal of the Royal Statistical Society: Series B}, year={1977}, volume={39}, number={1}, pages={1--38}}

@book{McLachlanPeel2000,
 author={McLachlan, Geoffrey J. and Peel, David}, title={Finite Mixture Models}, publisher={Wiley}, address={New York}, year={2000}}

@book{TitteringtonSmithMakov1985,
 author={Titterington, D. Michael and Smith, Adrian F. M. and Makov, Udi E.}, title={Statistical Analysis of Finite Mixture Distributions}, publisher={Wiley}, address={Chichester}, year={1985}}

@techreport{GhahramaniHinton1997,
 author={Ghahramani, Zoubin and Hinton, Geoffrey E.}, title={The {EM} Algorithm for Mixtures of Factor Analyzers}, institution={University of Toronto}, number={CRG-TR-96-1}, year={1996}, note={Revised February 1997}}

@article{McLachlanPeelBean2003,
 author={McLachlan, Geoffrey J. and Peel, David and Bean, Richard W.}, title={Modelling high-dimensional data by mixtures of factor analyzers},journal={Computational Statistics \& Data Analysis},year={2003},volume={41},number={3--4},pages={379--388}}

@article{BaekMcLachlanFlack2010,
 author={Baek, Jangsun and McLachlan, Geoffrey J. and Flack, Lloyd K.},title={Mixtures of factor analyzers with common factor loadings: Applications to the clustering and visualization of high-dimensional data},journal={IEEE Transactions on Pattern Analysis and Machine Intelligence},year={2010},volume={32},number={7},pages={1298--1309}}

@article{CeleuxGovaert1995,
 author={Celeux, Gilles and Govaert, G{\'e}rard},title={{G}aussian parsimonious clustering models},journal={Pattern Recognition},year={1995},volume={28},number={5},pages={781--793}}

@article{FraleyRaftery2002,
 author={Fraley, Chris and Raftery, Adrian E.}, title={Model-based clustering, discriminant analysis, and density estimation}, journal={Journal of the American Statistical Association},year={2002},volume={97},number={458},pages={611--631}}

@article{ViroliMcLachlan2019,
 author={Viroli, Cinzia and McLachlan, Geoffrey J.},title={Deep {G}aussian mixture models},journal={Statistics and Computing},year={2019},volume={29},number={1},pages={43--51}}

@inproceedings{TangSalakhutdinovHinton2012,
 author={Tang, Yichuan and Salakhutdinov, Ruslan and Hinton, Geoffrey E.},title={Deep mixtures of factor analysers},booktitle={Proceedings of the 29th International Conference on Machine Learning},year={2012},pages={505--512}}

@article{KockKleinNott2022,
 author={Kock, Lucas and Klein, Nadja and Nott, David J.},title={Variational inference and sparsity in high-dimensional deep {G}aussian mixture models},journal={Statistics and Computing},year={2022},volume={32}}

@article{PeelMcLachlan2000,
 author={Peel, David and McLachlan, Geoffrey J.},title={Robust mixture modelling using the $t$ distribution},journal={Statistics and Computing},year={2000},volume={10},number={4},pages={339--348}}

@article{LangeLittleTaylor1989,
 author={Lange, Kenneth L. and Little, Roderick J. A. and Taylor, Jeremy M. G.},title={Robust statistical modeling using the $t$ distribution},journal={Journal of the American Statistical Association},year={1989},volume={84},number={408},pages={881--896}}

@article{AndrewsMallows1974,
 author={Andrews, David F. and Mallows, Colin L.},title={Scale mixtures of normal distributions},journal={Journal of the Royal Statistical Society: Series B},year={1974},volume={36},number={1},pages={99--102}}

@article{WuMcLachlan2026Robust,
  title={Robust Deep Mixture Models},
  author={Wu, Jinran and McLachlan, Geoffrey J},
  journal={arXiv preprint arXiv:2608.01613},
  year={2026}
}

@article{WuWangMcLachlan2026SkewT,
  title={Deep Skew-t Mixture Models},
  author={Wu, Jinran and Wang, You-Gan and McLachlan, Geoffrey J},
  journal={arXiv preprint arXiv:2609.00773},
  year={2026}
}

@article{AzzaliniCapitanio2003,
 author={Azzalini, Adelchi and Capitanio, Antonella},title={Distributions generated by perturbation of symmetry with emphasis on a multivariate skew $t$-distribution},journal={Journal of the Royal Statistical Society: Series B},year={2003},volume={65},number={2},pages={367--389}}

@article{LeeMcLachlan2014,
 author={Lee, Sharon X. and McLachlan, Geoffrey J.},title={Finite mixtures of multivariate skew $t$-distributions: Some recent and new results},journal={Statistics and Computing},year={2014},volume={24},pages={181--202}}

@article{BrowneMcNicholas2015,
 author={Browne, Ryan P. and McNicholas, Paul D.},title={A mixture of generalized hyperbolic distributions},journal={Canadian Journal of Statistics},year={2015},volume={43},number={2},pages={176--198}}

@article{FuchsPommeretViroli2022,
 author={Fuchs, Robin and Pommeret, Denys and Viroli, Cinzia},title={Mixed deep {G}aussian mixture model: A clustering model for mixed datasets},journal={Advances in Data Analysis and Classification},year={2022},volume={16},number={1},pages={31--53}}

@article{MoustakiKnott2000,
 author={Moustaki, Irini and Knott, Martin},title={Generalized latent trait models},journal={Psychometrika},year={2000},volume={65},number={3},pages={391--411}}

@article{GolliniMurphy2014,
 author={Gollini, Isabella and Murphy, Thomas Brendan},title={Mixture of latent trait analyzers for model-based clustering of categorical data},journal={Statistics and Computing},year={2014},volume={24},number={4},pages={569--588}}

@article{McParlandGormley2016,
 author={McParland, Damien and Gormley, Isobel Claire},title={Model based clustering for mixed data: {clustMD}},journal={Advances in Data Analysis and Classification},year={2016},volume={10},number={2},pages={155--169}}

@book{McCullaghNelder1989,
 author={McCullagh, Peter and Nelder, John A.},title={Generalized Linear Models},edition={2},publisher={Chapman \& Hall},address={London},year={1989}}

@article{WeiTanner1990,
 author={Wei, Greg C. G. and Tanner, Martin A.},title={A {Monte Carlo} implementation of the {EM} algorithm and the poor man's data augmentation algorithms},journal={Journal of the American Statistical Association},year={1990},volume={85},number={411},pages={699--704}}

@article{Nielsen2000,
 author={Nielsen, S{\o}ren F.},title={The stochastic {EM} algorithm: Estimation and asymptotic results},journal={Bernoulli},year={2000},volume={6},number={3},pages={457--489}}

@article{Schwarz1978,
 author={Schwarz, Gideon},title={Estimating the dimension of a model},journal={The Annals of Statistics},year={1978},volume={6},number={2},pages={461--464}}

@article{HubertArabie1985,
 author={Hubert, Lawrence and Arabie, Phipps},title={Comparing partitions},journal={Journal of Classification},year={1985},volume={2},pages={193--218},}

@article{mclachlan2019finite,
  title={Finite mixture models},
  author={McLachlan, Geoffrey J and Lee, Sharon X and Rathnayake, Suren I},
  journal={Annual Review of Statistics and its Application},
  volume={6},
  number={1},
  pages={355--378},
  year={2019},
  publisher={Annual Reviews}
}

@book{mclachlan2008algorithm,
  title={The EM Algorithm and Extensions},
  author={McLachlan, Geoffrey J and Krishnan, Thriyambakam},
  year={2008},
  address={New York},
  publisher={John Wiley \& Sons}
}

@article{Gower1971,
  author={Gower, John C.},
  title={A general coefficient of similarity and some of its properties},
  journal={Biometrics}, year={1971}, volume={27}, number={4}, pages={857--871}
}

@article{Rousseeuw1987,
  author={Rousseeuw, Peter J.},
  title={Silhouettes: A graphical aid to the interpretation and validation of cluster analysis},
  journal={Journal of Computational and Applied Mathematics}, year={1987}, volume={20}, pages={53--65}
}

@inproceedings{SelosseEtAl2020,
  author={Selosse, Margot and Gormley, Claire and Jacques, Julien and Biernacki, Christophe},
  title={A bumpy journey: Exploring deep {Gaussian} mixture models},
  booktitle={I Can't Believe It's Not Better Workshop at NeurIPS}, year={2020}
}

@misc{MixedDeepMM2026,
  author={Wu, Jinran and McLachlan, Geoffrey J.},
  title={{MixedDeepMM}: Deep Mixture Models for Continuous and Mixed-Type Data},
  year={2026}, note={R package version 0.1.0},
  url={https://cran.r-project.org/package=MixedDeepMM}
}
\end{document}